\documentclass[trackchanges, twocolumn]{aastex701}

\defcitealias{Sarr25}{Sarrouh \& Asada et al. (2026)}
\defcitealias{FFSparkle}{L. Mowla \& K. Iyer et al. (2024)}

\begin{document}

\title{A Supermagnified Little Red Dot with a Resolved Host}

\author[orcid=0009-0000-2101-1938]{Jon Jude\v{z}}
\affiliation{University of Ljubljana, Faculty of Mathematics and Physics, Jadranska ulica 19, SI-1000 Ljubljana, Slovenia}
\email[show]{jon.judez@fmf.uni-lj.si}  

\author[orcid=0000-0002-9909-3491]{Roberta Tripodi}
\affiliation{INAF - Osservatorio Astronomico di Roma, Via Frascati 33, Monte Porzio Catone, 00078, Italy}
\email{roberta.tripodi@inaf.it}  

\author[0000-0003-3243-9969]{Nicholas S. Martis}
\email{nicholas.martis@fmf.uni-lj.si}
\affiliation{University of Ljubljana, Faculty of Mathematics and Physics, Jadranska ulica 19, SI-1000 Ljubljana, Slovenia}

\author[0000-0001-5984-0395]{Maru\v{s}a Brada\v{c}}
\email{marusa.bradac@fmf.uni-lj.si}
\affiliation{University of Ljubljana, Faculty of Mathematics and Physics, Jadranska ulica 19, SI-1000 Ljubljana, Slovenia}
\affiliation{Department of Physics and Astronomy, University of California Davis, 1 Shields Avenue, Davis, CA 95616, USA}

\author[0000-0001-8115-5845]{Rosa M. M\'erida}
\email{rosa.meridagonzalez@smu.ca}
\affiliation{Department of Astronomy and Physics and Institute for Computational Astrophysics, Saint Mary's University, 923 Robie Street, Halifax, Nova Scotia B3H 3C3, Canada}

\author[0000-0002-4201-7367]{Chris J. Willott}
\email{chris.willott@nrc.ca}
\affiliation{National Research Council of Canada, Herzberg Astronomy \& Astrophysics Research Centre, 5071 West Saanich Road, Victoria, BC, V9E 2E7, Canada}

\author[0000-0001-8325-1742]{Guillaume Desprez}
\email{guillaume.desprez@protonmail.com}
\affiliation{Kapteyn Astronomical Institute, University of Groningen, P.O. Box 800, 9700AV Groningen, The Netherlands}

\author[orcid=0000-0002-3364-4502]{Vesna Pirc Jev\v{s}enak}
\affiliation{University of Ljubljana, Faculty of Mathematics and Physics, Jadranska ulica 19, SI-1000 Ljubljana, Slovenia}
\email{vesna.pirc@fmf.uni-lj.si}  

\author[orcid=0000-0003-3983-5438]{Yoshihisa Asada}
\affiliation{Dunlap Institute for Astronomy and Astrophysics, 50 St. George Street, Toronto, Ontario, M5S 3H4, Canada}
\email{yoshi.asada@utoronto.ca}  

\author[orcid=0009-0009-4388-898X]{Gregor Rihtar\v{s}i\v{c}}
\affiliation{University of Ljubljana, Faculty of Mathematics and Physics, Jadranska ulica 19, SI-1000 Ljubljana, Slovenia}
\email{gregor.rihtarsic@fmf.uni-lj.si}  

\author[orcid=0000-0002-8434-880X]{Lilan Yang}
\affiliation{Department of Physics, School of Physics and Electronics, Hunan Normal University, Changsha 410081, People’s Republic of China}
\email{lxysps@rit.edu}  

\author[0000-0002-4542-921X]{Roberto Abraham}
\email{roberto.abraham@utoronto.ca}
\affiliation{David A. Dunlap Department of Astronomy and Astrophysics, University of Toronto, 50 St. George Street, Toronto, Ontario, M5S 3H4, Canada}
\affiliation{Dunlap Institute for Astronomy and Astrophysics, 50 St. George Street, Toronto, Ontario, M5S 3H4, Canada}

\author[0000-0003-2680-005X]{Gabriel B. Brammer}
\email{gabriel.brammer@nbi.ku.dk}
\affiliation{Niels Bohr Institute, University of Copenhagen, Jagtvej 128, DK-2200 Copenhagen N, Denmark}
\affiliation{Cosmic Dawn Center (DAWN), Jagtvej 128, DK-2200 Copenhagen N, Denmark}

\author[orcid=0000-0001-8489-2349]{Viicente Estrada-Carpenter}
\affiliation{School of Earth and Space Exploration, Arizona State University, Tempe, AZ 85287, USA}
\affiliation{Beus Center for Cosmic Foundations, Arizona State University, Tempe, AZ 85287, USA}
\email{vestrad9@asu.edu} 

\author[0000-0002-7547-3385]{Jasleen Matharu}
\affiliation{Cosmic Dawn Center (DAWN), Jagtvej 128, DK-2200 Copenhagen N, Denmark}
\affiliation{Niels Bohr Institute, University of Copenhagen, Jagtvej 128, DK-2200 Copenhagen N, Denmark}
\email{jamatharu@mpia.de}

\author[0000-0002-8530-9765]{Lamiya A. Mowla}
\email{lmowla@wellesley.edu}
\affiliation{Whitin Observatory, Department of Physics and Astronomy, Wellesley College, 106 Central Street, Wellesley, MA 02481, USA}
\affiliation{Center for Astronomy, Space Science, and Astrophysics, Independent University Bangladesh, Dhaka, Bangladesh}

\author[0000-0002-9330-9108]{Adam Muzzin}
\email{muzzin@yorku.ca}
\affiliation{Department of Physics and Astronomy, York University, 4700 Keele St., Toronto, Ontario, M3J 1P3, Canada}

\author{Ga\"el Noirot}
\email{gnoirot@stsci.edu}
\affiliation{Space Telescope Science Institute, 3700 San Martin Drive, Baltimore, Maryland 21218, USA}

\author[0000-0001-8830-2166]{Ghassan T. E. Sarrouh}
\email{gsarrouh@yorku.ca}
\affiliation{Department of Physics and Astronomy, York University, 4700 Keele St., Toronto, Ontario, M3J 1P3, Canada}

\author[0000-0002-7712-7857]{Marcin Sawicki}
\email{marcin.sawicki@smu.ca}
\affiliation{Department of Astronomy and Physics and Institute for Computational Astrophysics, Saint Mary's University, 923 Robie Street, Halifax, Nova Scotia B3H 3C3, Canada}

\author[0009-0000-8716-7695]{Sunna Withers}
\email{sunnaw@yorku.ca}
\affiliation{Department of Physics and Astronomy, York University, 4700 Keele St., Toronto, Ontario, M3J 1P3, Canada}

\author[0000-0002-0243-6575]{Jacqueline Antwi-Danso}
\email{j.antwidanso@utoronto.ca}
\affiliation{David A. Dunlap Department of Astronomy and Astrophysics, University of Toronto, 50 St. George Street, Toronto, Ontario, M5S 3H4, Canada}

\author[0009-0001-0778-9038]{Giordano Felicioni}
\affiliation{University of Ljubljana, Faculty of Mathematics and Physics, Jadranska ulica 19, SI-1000 Ljubljana, Slovenia}
\email{giordano.felicioni@fmf.uni-lj.si}

\author[orcid=0000-0001-9414-6382]{Anishya Harshan}
\affiliation{Kavli Institute for Cosmology, University of Cambridge, Madingley Road, Cambridge, CB3 0HA, United Kingdom}
\affiliation{Cavendish Laboratory - Astrophysics Group, University of Cambridge, 19 JJ Thomson Avenue, Cambridge, CB3 0HE, United Kingdom}
\email{at2207@cam.ac.uk}  

\author[orcid=0000-0002-5694-6124]{Vladan Markov}
\affiliation{University of Ljubljana, Faculty of Mathematics and Physics, Jadranska ulica 19, SI-1000 Ljubljana, Slovenia}
\email{vladan.markov@fmf.uni-lj.si}  

\author[orcid=0009-0000-5385-8674]{Maya Merchant}
\affiliation{Department of Physics and Astronomy, York University, 4700 Keele St., Toronto, Ontario, M3J 1P3, Canada}
\email{merchm@yorku.ca}  

\author[0000-0003-0780-9526]{Visal Sok}
\email{visal.sok@colorado.edu}
\affiliation{Department of Astrophysical and Planetary Sciences, University of Colorado, 2000 Colorado Ave, Boulder, CO 80309, USA}

\begin{abstract}

Little red dots (LRDs) are an intriguing new class of sources in the early universe discovered by the James Webb Space Telescope (JWST) characterized by V-shaped spectra and extremely compact sizes. Their underlying mechanisms still remain the subject of ongoing debate. We present JWST observations of a gravitationally lensed LRD nicknamed ``Best Friend'' at  $z_{\rm spec} = 8.297\pm0.002$. Its high magnification factor of $\mu = 36^{+8}_{-12}$ makes it the highest magnified LRD reported to date and unveils both the resolved structure of its host and a central engine less than $7\,\mathrm{pc}$ in size. The spectrum of the central engine, in which broad lines are not detected, possibly owing to the limited S/N and spectral resolution, is better interpreted as black hole activity rather than a dense star cluster. The latter cannot be fully excluded, but if true, the resulting lower limit on stellar density would be $>1.25^{+0.4}_{-0.3}\cdot10^6\ \mathrm{M_{\odot}/pc}^2$, which is around 2 orders of magnitude higher than stellar densities of typical clusters at high-z. By performing a spatially resolved analysis of NIRCam photometry and NIRSpec/PRISM spectroscopy, we find that the host stellar mass is $\log(M_\star/M_{\odot}) = 7.3^{+0.2}_{-0.8}$. Best Friend demonstrates the power of lensing to reveal the morphologies of LRDs.

\end{abstract}



\section{Introduction} 

Little Red Dots (LRDs) are very compact sources at high redshifts characterized by a distinct V-shaped spectrum with a blue slope in rest-frame ultra violet (UV) and red slope in rest-frame optical \citep{Kocevski_criteria}, with the transition consistently occurring around the Balmer limit (e.g.,  \citealt{Setton_turnover}). They are notably faint in X-rays (e.g., \citealt{Maiolino_xrays, AnannaXray}), radio waves (e.g., \citealt{RadioSilence}), and lack variability (e.g., \citealt{KokuboVariability}), while frequently showing broad Balmer and Paschen lines (e.g., \citealt{hviding25, deGraafLRDs, perezgonzalez26}).

Many of them are showing broad emission lines (e.g., \citealt{hviding25}), hence a purely stellar nature for all LRDs is unlikely. Furthermore, the stellar masses required to reproduce LRD properties would be too large causing unphysical stellar densities \citep{Baggen-only-stellar} and implying X-ray emission \citep{deGraafRubyBalmerBreaks}. Some LRDs also exhibit deeper Balmer breaks than what is allowed by stellar physics \citep{deGraafRubyBalmerBreaks} and at least 20\% of the spectroscopically confirmed LRDs show absorption in the Balmer lines \citep{Inayohi_firstBHstar}. A dust-reddened Active Galactic Nucleus (AGN) scenario, implying lower stellar masses \citep{Matthee24, Labbe-LRDsample, Greene24AGNInterpretation} is most commonly adopted. This interpretation, however, faces several challenges, among them the many non-detections of LRDs with ALMA \citep{CaseyDust, AkinsALMA, SettonLackofDust} and MIRI faintness \citep{Williams24}, implying an apparent lack of strong dusty torus emission. A competing solution pictures LRDs with a dense gaseous cocoon or envelope surrounding the central black hole (e.g., \citealp{Inayohi_firstBHstar, NaiduBH*original}). The nature of LRDs remains a matter of active debate \citep{Inayoshi_critical}.

Keys to understanding the galaxy-black hole coevolution are the black hole (BH) to stellar mass ratios observed in LRDs. They are usually extremely high (e.g., \citealt{RobertaNatCo, JonesBHmSm, DurodolaBHmSm, Juodžbalis}), however, there are large uncertainties in the measured BH masses that are almost universally estimated from single-epoch measurements of Balmer line widths \citep{Greene&HoRelation} and assumptions of virial motion, but see \citet{Juodžbalis}. \citet{GuptaMBHMstar} summarize that the observed ratios at $z>4.5$ surpass the well constrained local relation significantly ($\gtrsim1$ dex). This can point to very early BH mass growth (e.g., \citealt{KingFastSMBHgrowth}) or BH formation from heavy seeds (e.g., \citealt{Pacucci25MassiveSeeds}). It can also indicate a lack of host-BH coevolution at early times, with host galaxies assembling their masses relatively more slowly. However, these observed elevated ratios can also partly be due to observational biases \citep{LiBHmSmSelectionBias} and lack of proper calibrations.

Equally importantly, stellar mass measurements of the host galaxies in AGN suffer from outshining and unknown continuum contributions from central engines. For LRDs the latter is particularly poorly understood (e.g., \citealt{Gentile-quasi_star_model}). Observationally LRDs mostly remain unresolved under normal circumstances, which makes them very susceptible to these effects as the light of the host galaxy is normally merged with the light of the central engine.

\begin{figure*}[t]
    \centering
    \includegraphics[width=0.438\linewidth, keepaspectratio]{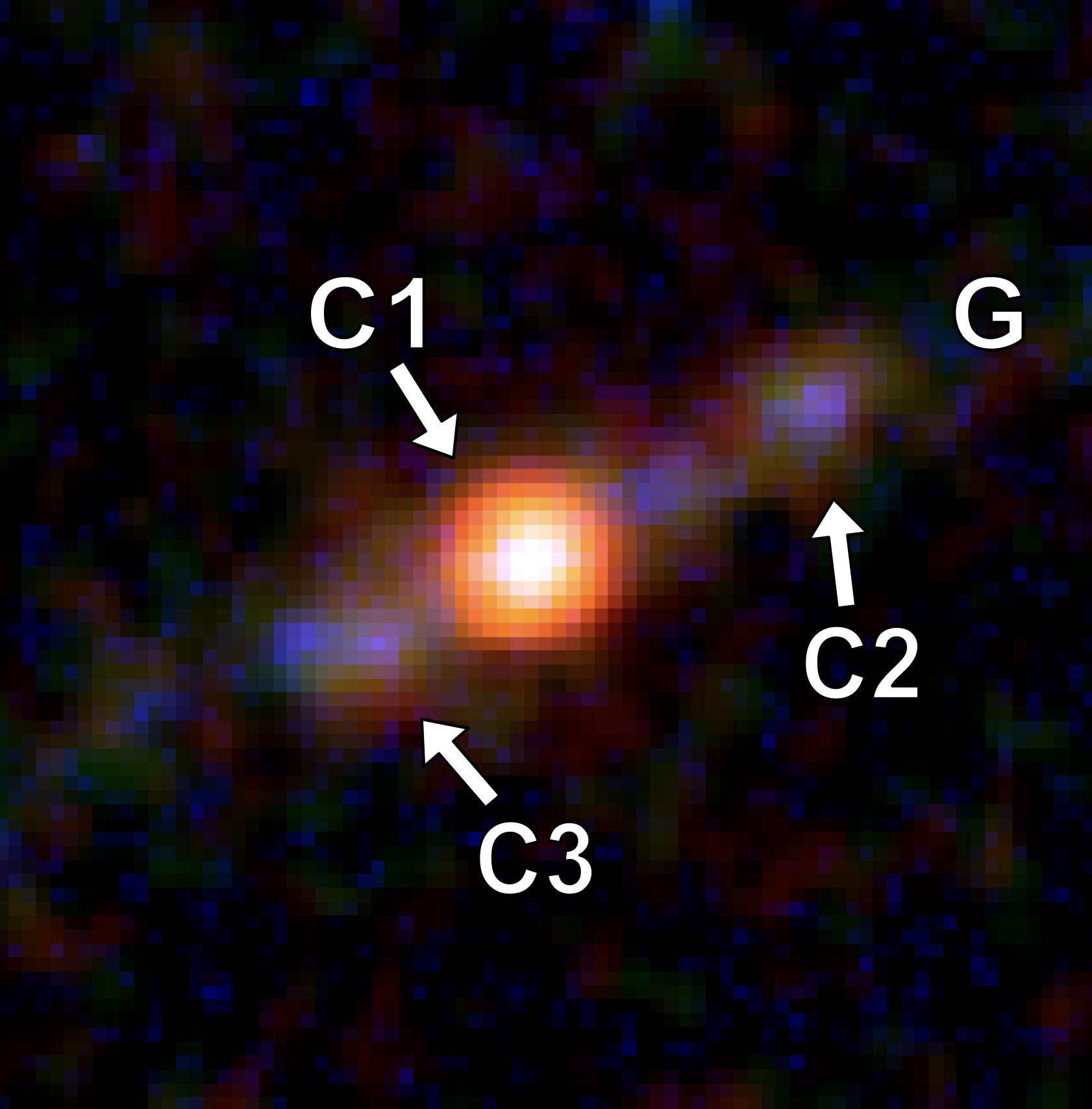}
    \hfill
    \raisebox{-0.08cm}{\includegraphics[width=0.542\linewidth, keepaspectratio]{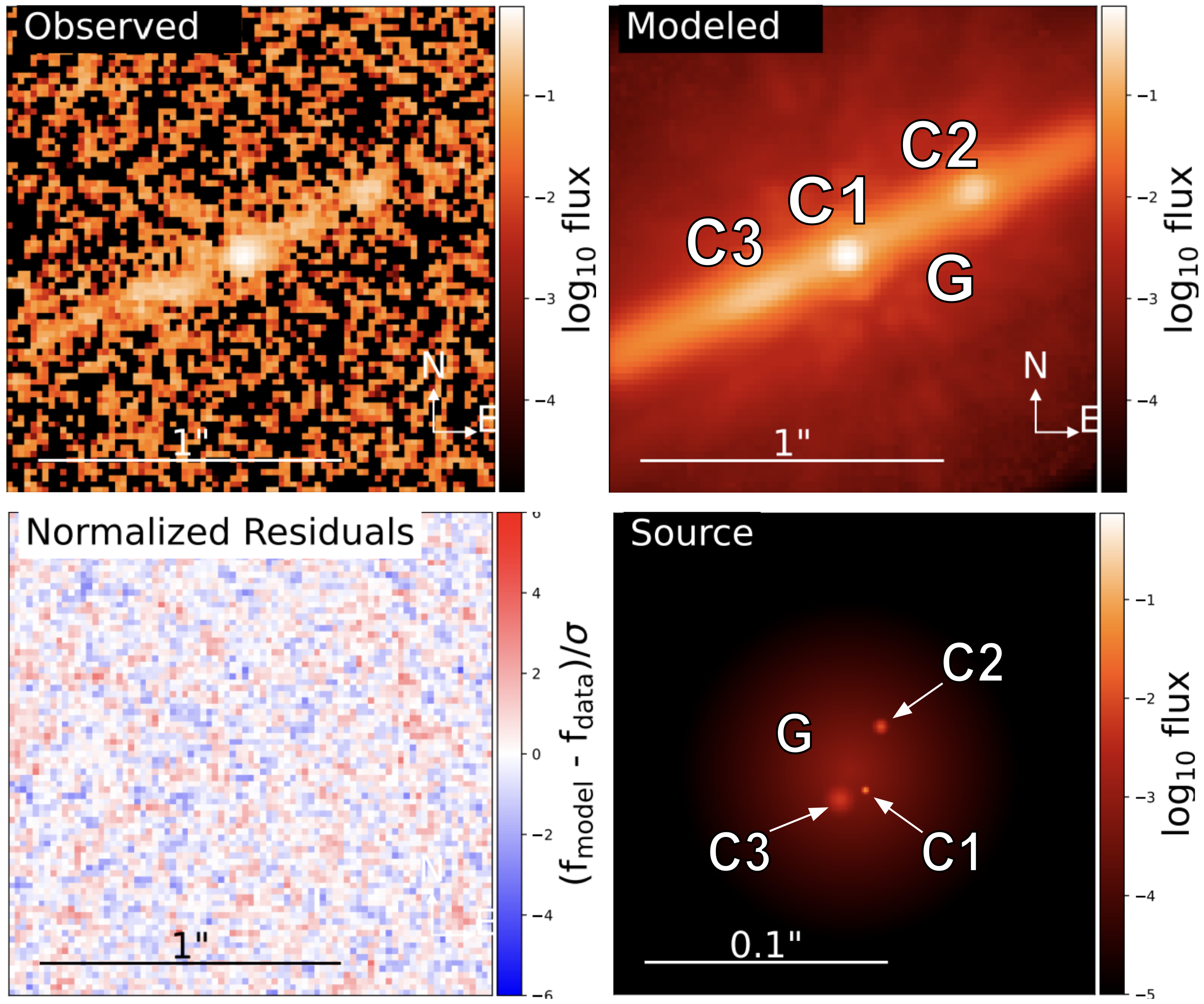}}
    \caption{\textbf{Left:} RGB of Best Friend. Seen is the underlying light of the galaxy (labeled G) and 3 clumps (labeled C1 to C3). Clump 1 is the brightest and has characteristics of an LRD. \textbf{Top middle:} F200W NIRCam image used for lensing reconstruction. \textbf{Top right:} Best modeled image. This is the result of forward modeling the four source components. \textbf{Bottom middle:} Normalized residuals. \textbf{Bottom right:} Best source-plane model after iterative forward modeling on the full set of lens model samples. Note that C1 is by far the brightest clump and is found to be located $50^{+10}_{-10}\,\mathrm{pc}$ from the galaxy center.}
    \label{fig:morphology}
\end{figure*}

The most promising way to solve this is by observing gravitationally lensed LRDs with resolved hosts. However, even in many known lensed LRDs, like the multiply imaged A2744-QSO1 \citep{FurtakQSO1Original}, the host galaxy remains undetected in imaging, demonstrating the need for very large magnifications. A few promising resolved lensed LRDs have been found lately. \citet{Yanagisawa_red_eyes} analyzed the photometry of a pair of LRDs that lie in a lensed galaxy dubbed ``Red Eyes'' at $z\sim7$ with magnification $\mu\sim20$. Another example is a doubly-imaged LRD in cluster A383 at $z=6.027$, analyzed independently in \citet{GolubchikA383} and \citet{BaggenA383} ($\mu\sim16$) showing a resolved neighbor rather than a host. Such objects are also opening questions on the effects of environment and the evolutionary paths of LRDs. For a detailed discussion, we refer to the work by \citet{RosiMeridaEnvironment} who also analyzed a rare non-lensed resolved LRD system nicknamed ``Stingray''.

In this paper we present a unique LRD dubbed ``Best Friend''. It lies at a redshift of  $z_{\rm spec} = 8.297\pm0.002$ in the field of galaxy cluster MACS J1423.8+2404 \citepalias{FFSparkle}, observed as part of the \textit{CAnadian NIRISS Unbiased Cluster Survey} (CANUCS, \citealt{Chris2022}). Its high magnification factor of $\mu = 36^{+8}_{-12}$ makes Best Friend the most highly magnified LRD discovered to date and shows a well resolved host which enables a study of its stellar properties.

Throughout this work, we assume a flat $\Lambda$CDM cosmology with $\Omega_\Lambda=0.7$, $\Omega_{\rm m}=0.3$, and $H_0=70~{\rm km\,s^{-1}\,Mpc^{-1}}$, AB magnitude system \citep{MagnitudesOke} and a Chabrier initial mass function \citep[IMF,][]{ChabrierIMF}.

\begin{deluxetable*}{rllll}
\tablewidth{0pt}
\tablecaption{Morphological forward modeling}
\tablehead{
\colhead{} & \colhead{Clump 1 (C1)} & \colhead{Underlying galaxy (G)} & \colhead{Clump 2 (C2)}& \colhead{Clump 3 (C3)}
}
\startdata
Size\tablenotemark{a} &
\begin{tabular}{c}
$<6.8\,\mathrm{pc}\ (2\sigma)$ \\
$<1.4\, \mathrm{mas}$
\end{tabular}
&
\begin{tabular}{c}
$120^{+40}_{-40}\,\mathrm{pc}$ \\
$25^{+8}_{-8}\, \mathrm{mas}$
\end{tabular}
&
\begin{tabular}{c}
$7^{+2}_{-4}\,\mathrm{pc}$ \\
$1.4^{+0.4}_{-0.7}\, \mathrm{mas}$
\end{tabular}
&
\begin{tabular}{c}
$14^{+5}_{-5}\,\mathrm{pc}$ \\
$3^{+1}_{-1}\, \mathrm{mas}$
\end{tabular}
\\ \tableline
Distance from the center
&
\begin{tabular}{c}
$50^{+10}_{-10}\,\mathrm{pc}$ \\
$10^{+2}_{-3}\, \mathrm{mas}$
\end{tabular}
&
defines center
& 
\begin{tabular}{c}
$115^{+20}_{-20}\,\mathrm{pc}$ \\
$24^{+5}_{-4}\, \mathrm{mas}$
\end{tabular}
&
\begin{tabular}{c}
$64^{+30}_{-30}\,\mathrm{pc}$ \\
$13^{+7}_{-6}\, \mathrm{mas}$
\end{tabular}
\enddata
\tablenotetext{a}{Clumps were modeled with Gaussian light profiles and the underlying galaxy with a Sérsic profile. Sizes are given as half-light radii.}
\tablecomments{Uncertainties are given as $1\sigma$ error bars, except for the size of the C1 which is given with a $2\sigma$ upper limit.}
\label{BigTable}
\end{deluxetable*}

\section{Data} \label{sec:Data}

JWST near-infrared imaging and spectroscopy of the Best Friend were obtained as part of CANUCS (PI Willott, PID: 1208, \citealt{Chris2022}). Details of the data acquisition and reduction along with the selected data products can be found in the CANUCS first data release \citetalias{Sarr25}, on the program website \url{https://niriss.github.io/} and on MAST (DOI: \dataset[10.17909/ph4n-6n76]{http://dx.doi.org/10.17909/ph4n-6n76}).

Cluster MACS J1423.8+2404 (MACS 1423, hereafter, at $z_{spec} = 0.5450\pm0.0005$) was observed in February 2023 with JWST Near Infrared Camera (NIRCam) filters F090W, F115W, F150W, F200W, F277W, F356W, F410M and F444W each with 6.4\,ks exposure time. This resulted in the corresponding 3-$\sigma$ depths of 29.51, 29.58, 29.73, 29.86, 30.12, 30.21, 29.50, 29.84 mag respectively, based on 0.3\,\arcsec aperture photometry. We subtracted bright cluster galaxies following the procedure described in Section 3 of \citet{Mart24BCGsub}. We also produced custom empirical point spread functions (PSFs) as outlined in \citet{Sarr24PSF}.

CANUCS revisited the same cluster 3 months later for spectroscopic follow-up with the JWST Near Infrared Spectrograph (NIRSpec) Micro-Shutter Assembly (MSA) with a low-resolution PRISM/CLEAR disperser ($R \sim 100$). Best Friend was observed in one 2.9\,ks MSA configuration. It was nodded over a 3-shutter slitlet, resulting in the total of 5 MSA shutters of data. Details of the initial spectra processing are given in \citet{Desp24}. Further details are also outlined in Sect. 3.3 of the CANUCS data release paper \citetalias{Sarr25}.

To obtain delensed (intrinsic) properties we use the cluster lens model described in \citetalias{FFSparkle}. It was built with the code \texttt{Lenstool} \citep{Jullo2007}. Lensing properties are produced for the best model and for the 100 randomly selected model sets from the optimization iteration process, to allow error estimation. Magnification for the Best Friend is $\mu = 36^{+8}_{-12}$ with a $2\sigma$ interval of $\mu = [21,63]$. The tangential magnification is $\mu_{tan} = 28^{+6}_{-8}$. The lens model does not predict any multiple images for Best Friend, because its predicted source-plane position approaches the caustic cusp but does not cross the caustic curve. Also, visually no multiple images are seen.

\section{Data Analysis} \label{sec:Data analysis}

\subsection{Morphology and Size Measurements} \label{subsec:Morphology & Size measurements}

The Best Friend is well resolved because of strong lensing magnification. It stretches $\sim 1\,\arcsec$ in length on the sky. To explore morphological parameters of the galaxy, we accounted for magnification and distortion of lensing. For this reconstruction we used the modified version of \texttt{Lenstruction} \citep{Lenstruction1,Lenstruction2,Lenstruction3}. \texttt{Lenstruction} uses a forward-modeling approach by creating 2D light profile models in the source plane and then applies the effects of lensing and the telescope's PSF. In an iterative way, it performs a Bayesian optimization of residuals between the modeled image and the data. In order to include the uncertainties of the cluster lens model, we used 100 lens model samples, which propagated magnification and deflection uncertainties into the morphological results.

We modeled Best Friend with 4 components (see Fig.~\ref{fig:morphology}). First a broad underlying component (labeled G) that we modeled with a 2D Sérsic light profile \citep{Sersic1986} with a Sérsic index of $n=1$ stretching throughout the galaxy to represent the diffuse light from the host galaxy. We also added 3 clumps (labeled C1 to C3) modeled with 2D Gaussian profiles. Clump 1 is the brightest and closest to the center of the galaxy and has a prominent red color. Clump 2 is also distinctly seen, whereas clump 3 is more extended and less distinguished from the underlying galaxy, however, modeling prefers the solution with all 3 clumps. Clumps 2 and 3 are blue and do not show any noticeable color difference with regards to the underlying galaxy light therefore we treat them as purely stellar and as part of the host. For the forward modeling, we selected the F200W NIRCam image, because of its relatively small PSF (0.066\,\arcsec), while still collecting substantial flux and effectively capturing both the blue and red components of the object.

\subsection{Spectroscopy} \label{subsec:Resolved spectroscopy}

\begin{figure*}
    \centering
    \includegraphics[width=\textwidth,
                 height=\dimexpr\textheight-9.2\baselineskip\relax,
                 keepaspectratio]{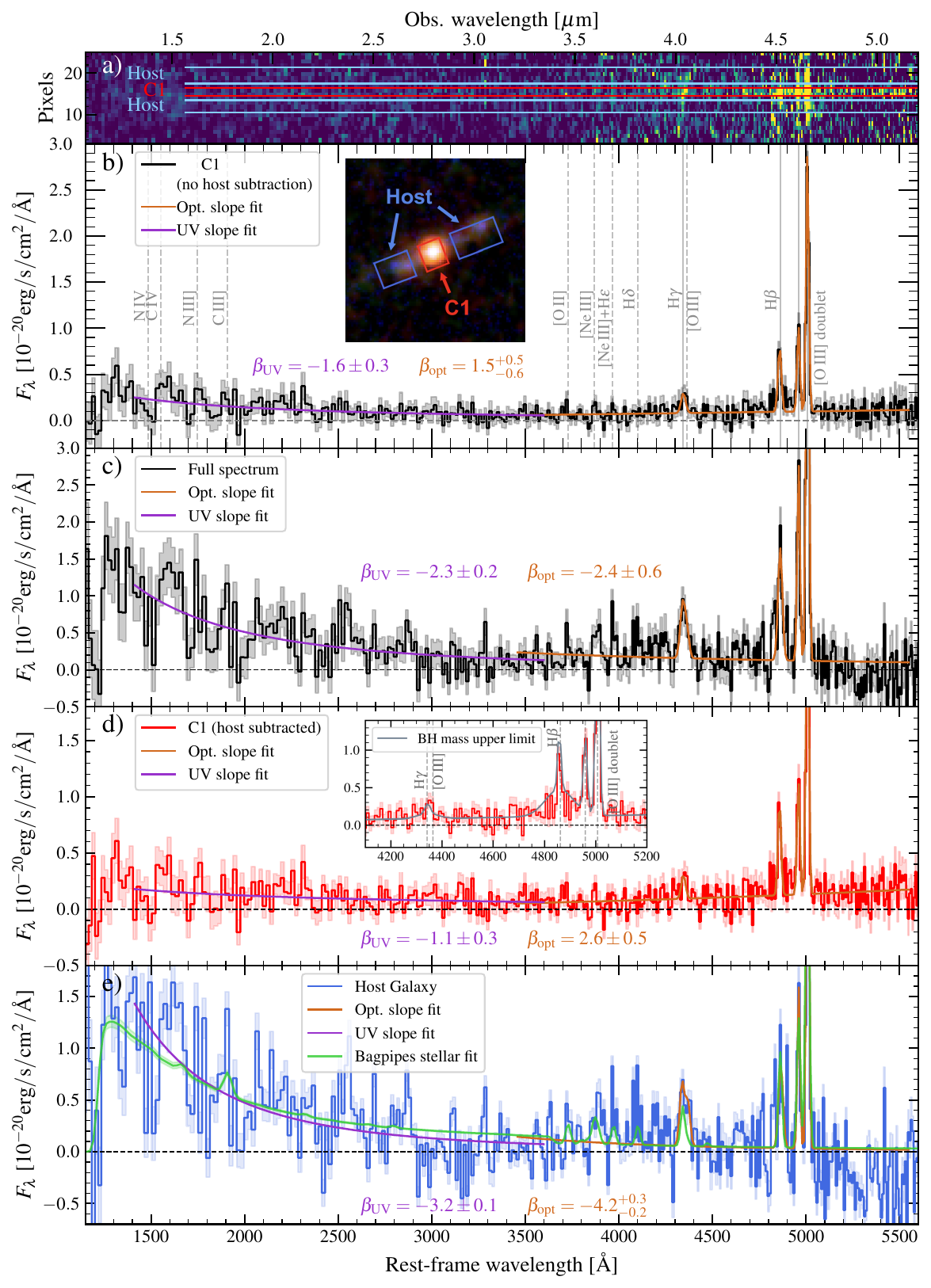}
    \caption{
        \textbf{a)} 2D spectrum with extraction bands.
        \textbf{b)} Spectrum of the C1 region without the host subtracted. In purple and gold are models for UV and rest optical respectively. Detected (nondetected) emission lines are marked with solid (dashed) lines. Added is also an RGB inset with blue (red) rectangles showing regions of host (C1) spectrum extraction. All spectra in this figure are observed (not corrected for magnification).
        \textbf{c)} Total spectrum of Best Friend.
        \textbf{d)} Spectrum of just the C1 with host galaxy subtracted. See text and Appendix~\ref{appsec: SpectroscopyExtractionAndNormalization} for details of normalization. Inset zooms in on the Balmer lines to show the model corresponding to the BH mass upper limit, $M_{BH}< 10^{8.9} \mathrm{M_\odot}$.
        \textbf{e)} Spectrum of only the host galaxy extracted from blue extraction regions in RGB inset and normalized. Galaxy model from \texttt{Bagpipes} with only stellar templates and \citet{DenseBasis} SFH is plotted in green.
    }
    \label{fig:spectra}
\end{figure*}

The MSA slit was well aligned with the local lensing shear direction. We took advantage of this by using the 2D spectrum to create our own custom spectroscopy extraction (see Fig.~\ref{fig:spectra}a). The clump 1 (C1) spectrum was obtained from a 2-pixel wide band (Fig.~\ref{fig:spectra}b). The host spectrum was taken from a 3-pixel wide band below the C1 and 4-pixel wide band above (Fig.~\ref{fig:spectra}e). We kept 1 pixel gaps between C1 and the host as buffer zones. These extraction windows were chosen based on imaging and are also seen in the inset of Fig.~\ref{fig:spectra}b. We also subtracted the underlying host spectrum from C1 (Fig.~\ref{fig:spectra}d). The full integrated spectrum of Best Friend was extracted from an 11-pixel wide band of the 2D spectrum effectively covering the signal from the entire length of the source (Fig.~\ref{fig:spectra}c). All spectra were normalized, corrected for any NIRSpec wavelength dependent effects and slitloss effects explained in detail in Appendix~\ref{appsec: SpectroscopyExtractionAndNormalization}. For normalizations, the resolved photometry was produced as described in Appendix~\ref{appsec: Resolved photometry} and summarized in Table~\ref{tab:sed}. Gaussian profile modeling of the Balmer lines in spectra does not detect broad components, though this cannot exclude them given PRISM resolution and shallow depth.

\subsection{Spectrophotometric SED Fitting} \label{subsec:DataAnalysisSEDfitting}

We used the SED fitting code \texttt{Bagpipes} \citep{2018Bagpipes, 2019Bagpipes} to infer parameters of the stellar populations, dust and possible AGN signatures in different components of Best Friend. We used the nonparametric star formation history (SFH) model from \citet{DenseBasis} but also performed separate runs with other SFH models like \citet{Leja19} and double power law to explore systematics. Modeling was done: 1) with purely stellar templates assuming no AGN, 2) with added standard AGN model from \citet{SloanAGNtemplates} which added a broken power law continuum with a break at Balmer limit $3646\,\mathrm{\AA}$ and broad $H_{\beta}$ and 3) with a \texttt{BagPipes} version modified by \citet{RosiMeridaModel, RosiCliff} that includes a Black body (BB) component, commonly used to analytically model the emission of the gas envelope within the BH* framework (e.g., \citealt{deGraafLRDs}), while also accounting for gas absorption at the Balmer limit, stellar, nebular emission and dust attenuation. See Appendix~\ref{appsec: DetailsOfSEDFitting} for details of spectrophotometric SED fitting.

\section{Results and Discussion} \label{sec:Results}

\subsection{Morphology} \label{subsec:Results-Morphology}

Results of morphological analysis with \texttt{Lenstruction} are summarized in Table \ref{BigTable}. Models of all 4 components were fitted for sizes (all reported as half-light radii) and positions. Clump 1 (C1) is unresolved, so we can measure the upper limit of its size which we find is only $<6.8\,\mathrm{pc}\ (2\sigma)$. Clump 2 is tentatively resolved at $7^{+2}_{-4}\,\mathrm{pc}$ and clump 3 is slightly larger at $14^{+5}_{-5}\,\mathrm{pc}$. The entire (host) galaxy has a size of $120^{+40}_{-40}\,\mathrm{pc}$. Interestingly, the C1 appears to be offset from the center by $50\pm10\,\mathrm{pc}$, though this might also be an effect of differential magnification.

\begin{figure*}
    \centering
    \includegraphics[width=0.8\textwidth]{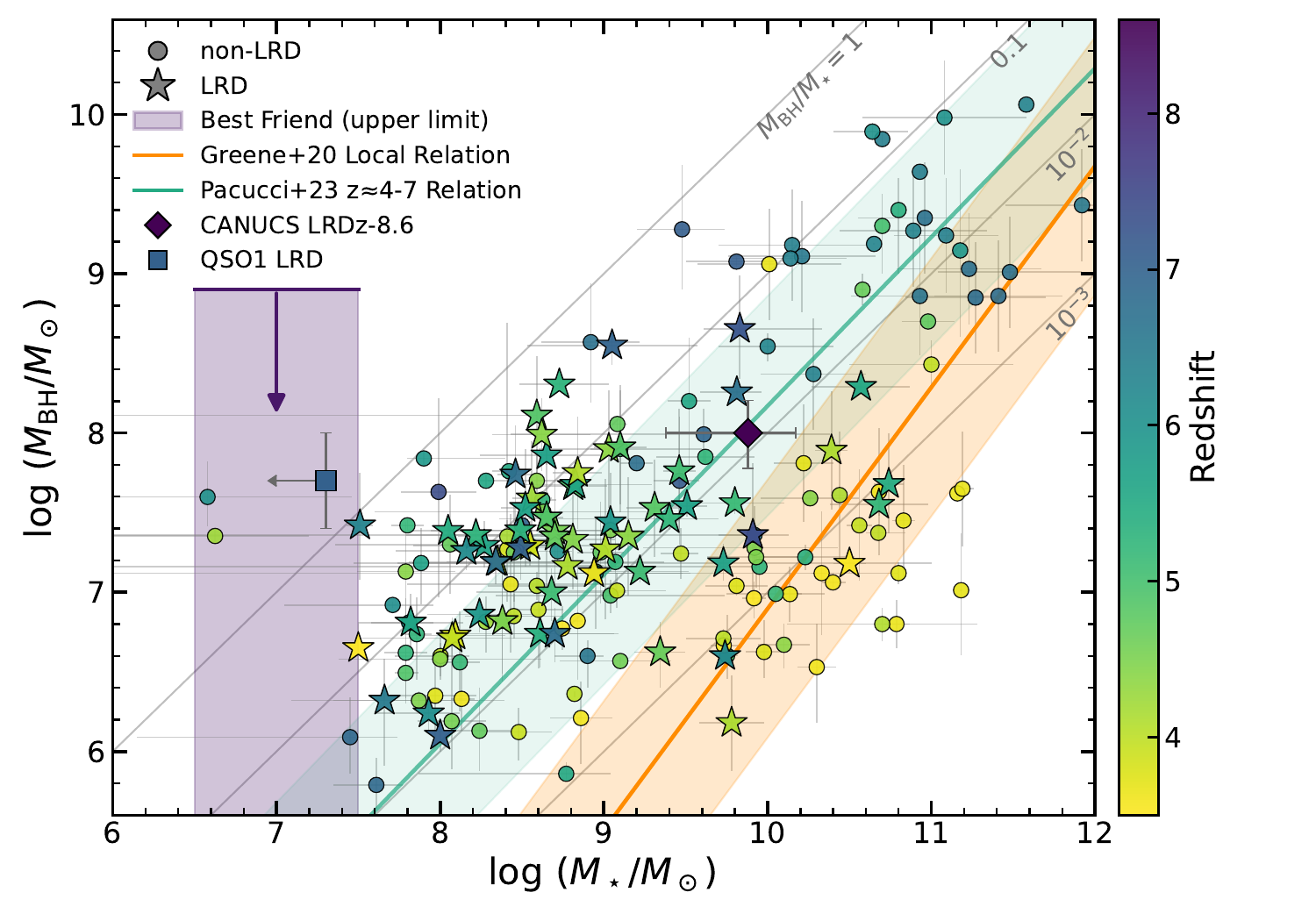}\\
    \caption{Black hole vs stellar mass relation of AGN objects spectroscopically observed by JWST above $z>3.5$ (\citet{GuptaMBHMstar}, modified) with redshift indicated by colors. We additionally show the position of CANUCS LRDz-8.6 \citep{RobertaNatCo} and QSO1 LRD \citep{Juodžbalis}. Best Friend is presented with a BH mass upper limit and the stellar mass interval that includes the systematic uncertainty of different SFH assumptions and lensing magnification uncertainty as a purple box. Shown are also the local universe relation by \citet{Greene} and relation at $z\sim4-7$  by \citet{Pacucci}. Lines for different $M_{BH}/M_{\star}$ are indicated.}
    \label{fig:BHvsS}
\end{figure*}

\subsection{UV and optical slopes} \label{subsec:Results-Slopes}

We measured the UV and optical slopes of spectra, with the transition at the Balmer limit ($3646\,\mathrm{\AA}$) (purple and gold lines with corresponding $\beta$ values in Fig.~\ref{fig:spectra}).
The spectrum of the central region without the host subtraction (C1 without the host subtraction) shows a strong V-shape with $\beta_{{\rm UV}} = -1.6\pm0.3$ and $\beta_{{\rm opt}} = 1.5^{+0.5}_{-0.6}$ satisfying the LRD definition \citep[see Fig.~\ref{fig:spectra}b,][]{Kocevski_criteria}. Additionally, we also fitted slopes on the spectrum of C1 with the host subtracted out (Fig.~\ref{fig:spectra}d). That spectrum is much redder, and the slope values are $\beta_{{\rm UV}} = -1.1\pm0.3$ and $\beta_{{\rm opt}} = 2.6\pm0.5$.

When we measure the same slopes on the total spectrum of Best Friend, we find that they are both blue with $\beta_{{\rm UV}} = -2.3\pm0.2$ and $\beta_{{\rm opt}} = -2.4\pm0.6$ (Fig.~\ref{fig:spectra}c). Therefore, because of the blue optical continuum, if the system was not lensed and its light was merged into a single point source, it would not pass the LRD criteria \citep{Kocevski_criteria}. It would be spectroscopically classified as a Little Blue Dot. Interestingly, this also appears to be true for the ``Red Eyes'' LRD system ($\mu \sim 20$ \citealt{Yanagisawa_red_eyes}). This suggests that we may be missing many LRDs embedded in the brighter light of their blue host galaxies. However, if the LRD in A383 by \citet{GolubchikA383} ($\mu \sim 16$) would be unlensed, it would still pass the selection. It is also theorized that a shallow optical slope can imply a transitional stage of an LRD as discussed in \citet{RosiMeridaEnvironment}.

\subsection{Stellar mass} \label{subsec:Results-StellarMass}

Spatially resolving the host galaxy enables the measurement of its stellar mass by fitting the light from only the stellar population without a direct contamination from the central engine continuum emission. \texttt{Bagpipes} with purely stellar templates was used. As per our definition of the host, the mass we report is comprised of the underlying host light (also the host underneath C1) and clumps C2 and C3. We find that the stellar mass is $\log(M_\star/M_{\odot}) = 7.3^{+0.2}_{-0.8}$, corrected for magnification (Fig.~\ref{fig:BHvsS}). The uncertainty, which also includes the magnification uncertainty, is mostly driven by differences in choices of SFHs and a young outshining population. Parameters used for this fit are shown in the ``Host'' column of Table~\ref{tab:sed}. More details and uncertainty discussions are provided in the Appendix~\ref{appsec: DetailsOfSEDFitting}.

\begin{deluxetable*}{lrccc}
\tablewidth{0pt}
\centering
\tablecaption{Photometry and spectrophotometric fitting results}
\tablehead{
  \colhead{} & \colhead{} & \colhead{Host} & \colhead{C1} & \colhead{Total}}
\startdata
a) & F115W (nJy) & $25^{+2}_{-2}$ & $4^{+1}_{-0.9}$ & $30^{+2}_{-2}$ \\
 & F150W (nJy) & $73^{+2}_{-2}$ & $14^{+1}_{-0.8}$ & $87^{+2}_{-2}$ \\
 & F200W (nJy) & $67^{+2}_{-2}$ & $15^{+0.7}_{-1}$ & $83^{+2}_{-2}$ \\
 & F277W (nJy) & $53^{+2}_{-2}$ & $22^{+1}_{-1}$ & $74^{+2}_{-2}$ \\
 & F356W (nJy) & $58^{+2}_{-2}$ & $42^{+2}_{-1}$ & $100^{+3}_{-2}$ \\
 & F410M (nJy) & $33^{+3}_{-3}$ & $98^{+2}_{-2}$ & $131^{+4}_{-4}$ \\
 & F444W (nJy) & $112^{+5}_{-3}$ & $182^{+3}_{-2}$ & $294^{+5}_{-4}$ \\
\hline
b) & $\beta_{{\rm UV}}$ &  $-3.2\pm0.1$ & $-1.1\pm0.3$ & $-2.3\pm0.2$ \\
 & $\beta_{{\rm opt}}$ & $-4.2^{{+0.3}}_{{-0.2}}$ & $2.6\pm0.5$ & $-2.4\pm0.6$\\
\hline
c) & $\log_{{10}}(M_*/\mu_{best} [M_{\odot}])$ & $7.30^{+0.02}_{-0.03}$ & $8.6^{+0.1}_{-0.1}$ & $7.77^{+0.07}_{-0.06}$ \\
 & $\log_{{10}}(SFR/\mu_{best} [M_{\odot}yr^{-1}])$ & $2.72^{+0.04}_{-0.10}$ & $2.4^{+0.2}_{-0.3}$ & $2.5^{+0.2}_{-0.2}$ \\
 & $A_V[mag]$ & $0.19^{+0.04}_{-0.04}$ & $2.7^{+0.3}_{-0.3}$ & $0.73^{+0.05}_{-0.05}$ \\
 & $t_{MW}$[Myr] & $470^{+10}_{-60}$ & $320^{+70}_{-70}$ & $380^{+50}_{-50}$
\enddata

\textbf{a)}~Photometry of respective components, not corrected for magnification (see Appendix~\ref{appsec: Resolved photometry}).
\textbf{b)}~Rest UV and optical slope values of spectra.
\textbf{c)}~\texttt{Bagpipes} runs results. Host and C1 are modeled with only stellar templates, Total BF also with an AGN model. Values for C1 are not values we think are likely to be true, since we believe the light we see is from an AGN/LRD. Shown are stellar mass, star formation rate (both corrected for magnification), dust attenuation and mass weighted age.

\label{tab:sed}
\end{deluxetable*}

We try fitting the light from just the central region (C1) with only stellar templates (see ``C1'' column of Table.~\ref{tab:sed}). In this case, large amounts of dust are invoked ($A_V = 2.4^{+0.2}_{-0.3}$) and very large stellar masses ($\log(M_\star/M_{\odot}) = 8.6^{+0.1}_{-0.1}$). The small size of the C1 region (conservatively using $2\sigma$ upper limit on size) would therefore mean extremely large effective stellar surface densities of $\Sigma_\star \geq 1.25^{+0.4}_{-0.3}\cdot10^6\ \mathrm{M_{\odot}/pc^2}$ which are around 2 orders of magnitude higher than stellar densities of typical stellar clusters at high-z at that size (e.g., \citealp{AdamoCosmicGems}) and exceed the densest known cores of local globular clusters (e.g., \citealt{MaxGCdensitied}). This demonstrates that a purely stellar explanation for the C1 is unlikely and that supermassive BH nature is preferred.

\subsection{Black hole properties} \label{subsec:Results-BHProperties}

Our spectroscopic fit does not favor any broad $H_{\beta}$ and $H_{\gamma}$ components; however, given the low resolution of the PRISM spectra ($R \sim 100$), this cannot fully rule out an AGN interpretation. By performing an MCMC fitting including a broad $H_{\beta}$ component, we derived an upper limit on the flux and line width of the $H_{\beta,\rm broad}$ that can still fit the data (inset of Fig.~\ref{fig:spectra}d fit). Adopting the relation from \citet{Greene&HoRelation} this gives us an upper limit on the BH mass of $M_{\rm BH} \leq 10^{8.9}\,\mathrm{M_{\odot}}$ (Fig.~\ref{fig:BHvsS}). However, we acknowledge the caveats of measuring BH masses from single-epoch Balmer line observations \citep{BerthemesBHmassProblematics}. We also checked how our spectrum normalization affected the fitting results by repeating the fit on the unnormalized version and found minimal effect on the resulting BH mass upper limit, well below the uncertainties.

When we fit the spectrum of C1 with host subtracted (and masked out emission lines) with modified \texttt{Bagpipes} of \citet{RosiMeridaModel}, we find that it is best fit with a combination of AGN and BB templates ($\chi^2_\nu = 1.64$), but also a purely black-body fit is consistent with the data ($\chi^2_\nu = 2.18$). Black body temperature is hard to estimate, because of the limited rest wavelength coverage towards the optical, but is found to be in the range between 4000 and 6000 K.

\section{Conclusions} \label{sec:Conclusions}

We present the spectroscopic and photometric investigation of a Little Red Dot with a resolved host galaxy called ``Best Friend''. It is the highest magnified LRD currently known ($\mu_{best} = 36^{+8}_{-12}$). High magnification and the resulting large apparent size of the host allow us to spatially separate the host galaxy from the central region and investigate them separately all while reaching spatial scales down to $\lesssim7\,\mathrm{pc}$. The central region (C1) has a clear V-shaped spectrum and so meets the LRD criteria \citep{Kocevski_criteria}, but because the available spectrum lacks resolution and sensitivity, we currently cannot confirm broad lines. Our work also shows that it is highly unlikely that this central region would be purely stellar in nature as this would imply extremely large stellar densities ($\Sigma_\star \geq 1.25^{+0.4}_{-0.3}\cdot10^6\ \mathrm{M_{\odot}/pc^2}$), two orders of magnitude higher than those observed in high-redshift stellar clusters \citep{AdamoCosmicGems} and higher than in the densest cores of local globular clusters \citep{MaxGCdensitied}. We find that the host stellar mass is $\log(M_\star/M_{\odot}) = 7.3^{+0.2}_{-0.8}$ with errorbars including lensing magnification uncertainty and estimates of systematic uncertainties from different SFH modeling prescriptions.

Without lensing magnification, the Best Friend would be a point source and the strong contribution of the host galaxy would make the spectrum blue in UV and optical, classifying it as a Little Blue Dot (e.g., \citealt{MadauUnification, perezgonzalez26}). Fig.~\ref{fig:BHvsS} shows that, thanks to the strong magnification, Best Friend has one of the lowest stellar masses discovered so far at high-z and, within the LRD population, it has one of the most robustly measured stellar masses thanks to its resolved host. At such high redshift, a high ratio of BH mass vs. stellar mass of $\lesssim40$ could imply a heavy-seed BH formation mechanism and challenge current host-BH coevolution models, but the BH mass needs to be measured accurately by deep higher resolution observations.

\begin{acknowledgments}
JJ, MB, GR, NM, AH, GF, RT and VM acknowledge support from the ERC Grant FIRSTLIGHT \#101053208, Slovenian national research agency ARIS through grants N1-0238 and P1-0188, and ESA PRODEX Experiment Arrangements No. 4000146646 and 4000149972. This research was also enabled by Canadian Space Agency grants 18JWST-GTO1 and 24JWGO3A12, and Natural Sciences and Engineering Research Council (NSERC) of Canada grants RGPIN-2020-06023, RGPAS-2020-00065, and RGPIN-2026-08047.  This research used the Canadian Advanced Network For Astronomy Research (CANFAR) operated in partnership by the Canadian Astronomy Data Centre and The Digital Research Alliance of Canada with support from the National Research Council of Canada the Canadian Space Agency, CANARIE and the Canadian Foundation for Innovation. The data were obtained from the Mikulski Archive for Space Telescopes at the Space Telescope Science Institute, which is operated by the Association of Universities for Research in Astronomy, Inc., under NASA contract NAS 5-03127 for JWST. Support for program \#4598 was provided by NASA through a grant from the Space Telescope Science Institute, which is operated by the Association of Universities for Research in Astronomy, Inc., under NASA contract NAS 5-03127.
\end{acknowledgments}

\section*{Data Availability}
All the JWST data used in this paper can be found on MAST: \dataset[10.17909/ph4n-6n76]{http://dx.doi.org/10.17909/ph4n-6n76}.

\begin{contribution}

JJ conducted morphological, photometric, and spectroscopic analysis, coordinated team contributions, and wrote and submitted the manuscript.
RT identified the object as an LRD candidate, fitted spectra for slopes and lines, and provided a BH mass estimate and assisted with interpretation.
NSM contributed to SED fitting, Galfit analysis, and initial image processing.
MB provided funding and mentorship and contributed to the manuscript and its interpretation.
RMM modified the Bagpipes modeling with BH* models and contributed to the interpretation.
CJW and YA contributed to the spectra analysis.
GD provided the cluster lens model and, together with GR, helped with lensing interpretation.
VPJ contributed to line modeling.
CJW, MB, AM, MS, NSM, GS, GD, YA, GR, RA, GB, VEC, JM, LM, GN and SW are CANUCS builders and provided survey design and data reduction.
All authors contributed to the manuscript.


\end{contribution}

%



\appendix

\section{Resolved photometry}
\label{appsec: Resolved photometry}

We measured photometry of the host galaxy and the photometry of C1. We used \texttt{Galfit} \citep{OriginalGalfit, NewestGalfit} to model fluxes in the image plane (see Figure \ref{fig:appendix_Galfit_figure} in Appendix). Clump 1 was modeled with a point source. We consider clumps 2 and 3 as purely stellar and treat them as being part of the host because they do not exhibit any noticeable color difference to the underlying galaxy. We modeled the underlying galaxy component with an elliptical 2D Sérsic light profile and clumps 2 and 3 with point sources. We found that we obtained more reliable photometry extractions if we used point source models for clumps 2 and 3 instead of small resolved models because the data lacks the constraining power in some filters to break degeneracies if we increase the number of parameters. Furthermore, fluxes from clumps 2 and 3 are directly summed with the underlying galaxy flux to obtain the host photometry, so the specific model used to reconstruct them individually does not affect our conclusions. This photometry of different components was used to correct spectra for slit losses as described in Appendix~\ref{appsec: SpectroscopyExtractionAndNormalization}.

If we estimated slopes using only wide-band photometry, we would recover a positive optical slope for the total object, selecting it as an LRD, but this is due to the large fraction of F444W flux in emission lines. With spectroscopy we found that the optical slope of the total object is negative, dropping it out of the LRD definition. This serves as a reminder that photometric selections of LRDs, although widely employed in the past, should be confirmed with spectroscopic data.

\begin{figure}[h]
    \centering
    \includegraphics[width=0.4\textwidth]{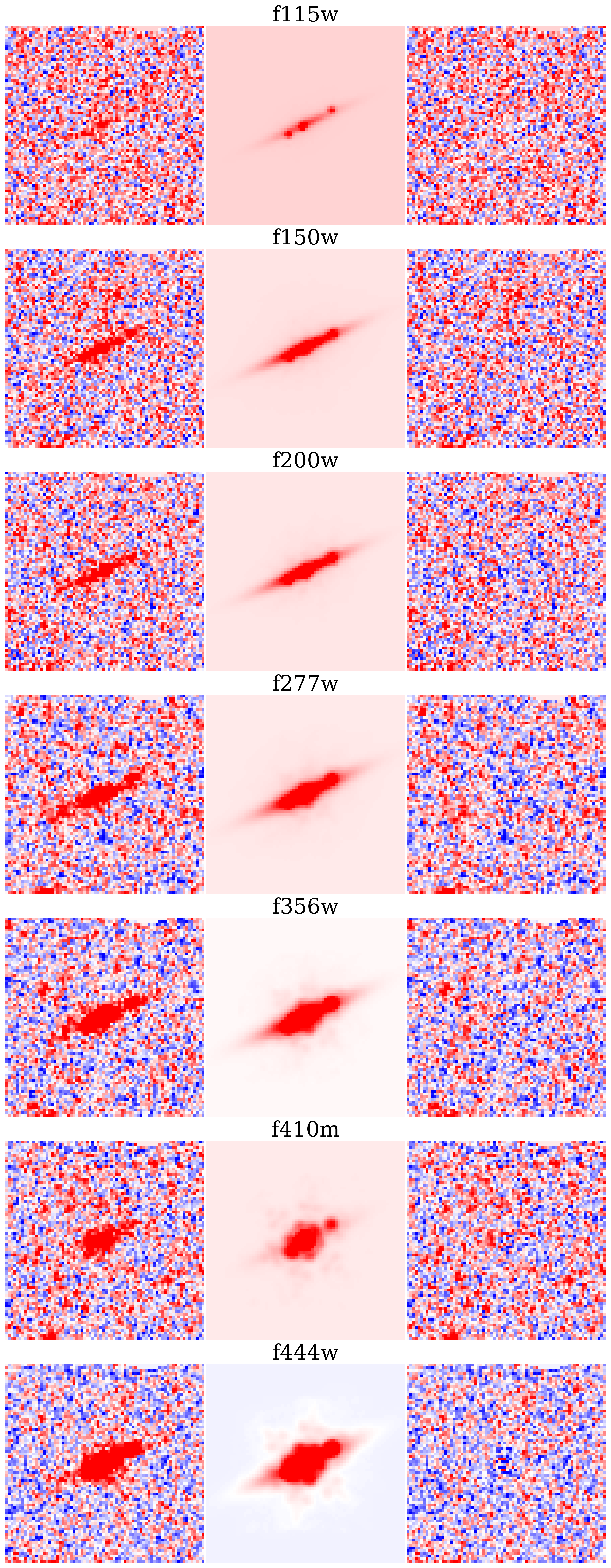}
    \caption{\texttt{Galfit} modeling of Best Friend. Each row shows the modeling in one filter. From left to right we plot images, best models and residuals. Modeling is performed with four components. 3 point sources for clumps and one elliptical Sérsic profile for the underlying galaxy.}
    \label{fig:appendix_Galfit_figure}
\end{figure}

\section{Spectroscopy extraction \& normalizations}
\label{appsec: SpectroscopyExtractionAndNormalization}

The object was captured in one MSA configuration. We subtracted the background using a master background, because the source filled most of the adjacent shutters used in the standard 3-point nod background subtraction method.

From the 2D spectrum, we extracted C1 from a central 2-pixel-wide aperture and the host from two flanking apertures, 3 pixels below and 4 pixels above C1. A 1-pixel gap between C1 and each host aperture was left as a buffer to limit cross-contamination (Fig.~\ref{fig:spectra}a). Each extracted spectrum was first normalized to the custom aperture photometry from the exact corresponding region where it was taken from (Fig.~\ref{fig:spectra}b inset) to correct for any wavelength-dependent effects of NIRSpec, like slit loss from different PSF sizes. The wavelength dependence was fit with a line, and correction slopes for the upper host slit, central slit, and lower host slit were $0.052/\mathrm{\mu m}$, $0.093/\mathrm{\mu m}$, and $-0.022/\mathrm{\mu m}$, respectively.

In the next step we subtracted the host spectrum from the spectrum of C1. This was done by normalizing accross different areas and assuming a constant surface brightness of the host galaxy in the observed image, which is a caveat, but there is no direct way of measuring how bright the host galaxy is underneath C1. In this way, we obtained the spectrum of just the C1, which is plotted in Fig.~\ref{fig:spectra}d.

Finally, we normalized the host spectrum and the C1-only spectrum to their respective photometric fluxes as found with \texttt{Galfit} modeling in F150W filter and described in Appendix~\ref{appsec: Resolved photometry}. This final normalization was done by wavelength-independent factors (constants), and its purpose was to account for all the light of the corresponding parts of the galaxy that did not land in the slits. F150W filter was used for this normalization because it lacks any strong emission lines and both the host and C1 are well detected in it. C1 spectrum was raised by a factor of $1.29$ and host by $2.07$. This correction also allowed us to account for the missing flux that fell in the gaps between shutters (MSA bars). In this way, we ensured that we measured the total stellar mass.

In addition to the extraction of specific parts of the galaxy, the total spectrum of the source was also obtained for comparison, and it was taken from an 11-pixel wide band in the 2D spectrum to gather the light from all parts of the object (Fig.~\ref{fig:spectra}c). It was corrected with the comparison to the total KRON photometry of the source with a first degree polynomial. This correction slope was $0.10/\mathrm{\mu m}$.

\section{Details of spectrophotometric SED fitting}
\label{appsec: DetailsOfSEDFitting}

We fixed the redshift to the spectroscopic redshift and chose the Chabrier IMF \citep{ChabrierIMF}. For dust we used the Calzetti dust attenuation curve \citep{CalzettiDust}, because its flatter slope is shown to be favored by high-z galaxies \citep{Markov25aDust, Shivaei26Dust, RobertaNatCo}. All stellar masses we report are corrected for magnification. 

Our final best estimate of the host stellar mass is $\log(M_\star/M_{\odot}) = 7.3^{+0.2}_{-0.8}$, including lensing uncertainty. This result accounts for the masses obtained using several different SFH models which produced substantially different results, almost by an order of magnitude in stellar mass and with different population ages. These differences are mostly driven by the fact that there is a recent elevated star-formation activity present with a star-formation rate of $\sim 10\,\mathrm{M_{\odot}/yr}$ which was detected across all models. This young population is outshining a possible older population and different SFH models treat outshining differently. Such differences were also demonstrated in \citet{VladanMarkov23}. If we used \citet{DenseBasis} SFH model, the \texttt{Bagpipes} result with purely stellar templates was $\log(M_\star/M_{\odot}) = 7.30^{+0.02}_{-0.03}$ and gives an old stellar population with mass weighted age of $470^{+10}_{-60}\,\mathrm{Myr}$. \citet{Leja19} SFH model, gives a much lower stellar mass at $\log(M_\star/M_{\odot}) = 6.45^{+0.08}_{-0.06}$ produced very recently with a mass weighted age of only $18^{+69}_{-5}\,\mathrm{Myr}$. When using the double power law SFH model, it gives an even lower mass weighted age of just $1.26^{+0.06}_{-0.02}\,\mathrm{Myr}$ with the mass of $\log(M_\star/M_{\odot}) = 6.34^{+0.03}_{-0.02}$.

We also performed spectrophotometric SED fitting of the full spectrum of Best Friend to see, how we would interpret it if it was not lensed and would be visible only as a point source. Results are summarized in the ``Total'' column of Table \ref{tab:sed}. This modeling uses Bagpipes setup with an AGN model that was set to have a break in the double power law at the Balmer limit.


\bibliography{bib}{}
\bibliographystyle{aasjournalv7}



\end{document}